\documentstyle[12pt]{article}
\begin{document}
\begin{titlepage}
\pagestyle{empty}
\setlength{\topmargin}{2 cm}
\setlength{\textheight}{21 cm}
\setlength{\textwidth}{16 cm}
\setlength{\oddsidemargin}{1.5 cm}
\topmargin -10mm
\begin{flushright}{\footnotesize Brown-HET-1058 \\

Oct 1996}
\end{flushright}
\date{}
\vskip 0.5cm
\begin{center}
{\Large \bf On the Equivalence between the Effective Potential and Zero-Point Energy}
\vskip 0.5cm
Jos\'e Alexandre Nogueira

{\it Departamento de F\'isica, Centro de Ci\^encias Exatas,}\\
{\it Universidade Federal do Espirito Santo,}\\
{\it 29.060-900 - Vit\'oria-ES - Brasil,}\\
{\it E-mail: } nogueira@cce.ufes.br\\
\vskip 0.5cm

Adolfo Maia Jr.

{\it Instituto de Matem\'atica, Universidade Estadual de Campinas,}\\
{\it 13.081-970 - Campinas-SP , Brasil}\\
{\it E-mail: } maia@ime.unicamp.br \\
{\it Department of Physics, Brown University,}\\
{\it 02912 Providence-RI , USA}
\end{center}

\begin{abstract}
We investigate a possible difference between the effective potential and zero-point energy. We define the zero-point ambiguity (ZPA) as the
difference between these two definitions of vacuum energy. Using the zeta function technique in order to obtain renormalized quantities,
 we show that ZPA
vanishes, implying that both of the above definitions of vacuum energy 
coincide for a large
class of geometries and a very general potential. In addition, we show 
explicitly that an extra term, obtained by E. Myers some years ago 
for the
ZPA, disappears when a scale parameter $\mu$ is consistently introduced in 
all zeta functions in order to keep them dimensionless.

\end{abstract}
\end{titlepage}
\newpage

\section{Introduction}

In modern quantum field theories, the concept and structure of the vacuum play a very important role. In quantum field theory, the vacuum is a well-defined quantum state, namely, the ground state of a system of fields. On the other hand, the evaluation of the vacuum energy gives rise to problems of divergence. In the case of free fields, in Minkowski space-time, the  vacuum energy is an infinite constant which has no physical meaning because we can get
rid of it consistently, by shifting the energy scale. However, vacuum fluctuations 
 have been shown to lead to observable effects, such as Lamb shift, Casimir effect, particle production in strong fields, etc.

Actually, there are several variations on the concept of vacuum energy. One of the most commonly used procedures to obtain the vacuum energy is the straightforward evaluation of an infinite sum over eigenvalues of zero-point energy. Although this sum has the intuitive and useful interpretation of vacuum energy, in many important problems one often uses the minimum of the effective potential, obtained from the functional approach to quantum field theory(QFT), as a series in loops (or, equivalently, in $\hbar$). The effective potential plays a central role in  theories with spontaneous symmetry breaking[1-7], compactification of the extra dimensions in Kaluza-Klein theory[13], inflationary universe models[14], etc.

In this paper we investigate a possible difference between the effective potential and zero-point energy. We name
this difference
as {\it zero-point ambiguity(ZPA)}.
Roughly speaking, the zero-point ambiguity arises because the evaluation of the effective potential, using the zeta function method, implicitly removes the divergences whereas, for the zero-point energy, an explicit renormalization
becomes necessary in order to get a finite result. Below, we
calculate explicitly the non-renormalized ZPA and show by using a suitable
prescription for renormalization (Salam-Strathdee[9]) that the renormalized ZPA 
vanishes, thus implying that the effective potential coincides with renormalized zero-point energy. In our previous work[12] this was shown, explicitly, for the particular case of Casimir plates with a general potential. In this work, our method is more general so that our results are valid for a large class of
geometries or, in other words, the fields we are considering are defined
on an arbitrary smooth submanifold of the Minkowski space-time.

Some years ago E. Myers showed[10] that vacuum energy calculated via the effective potential could be interpreted like zero-point energy except for an extra term. As we shall see this extra term arises when the scale parameter $\mu$ is not included in the definition of the generalized zeta function associated to the Hamiltonian operator for the sum over zero-point energy. Differently from the ZPA defined above, even after renormalization, Myers' extra term does not vanish and he conclude that for some theories, namely, for those in which, in our notation, $\zeta_{m}(0) \neq 0$, this simple identification between the two above definitions of vacuum energy breaks down. As noted by
Treml[11], this is a mistake since a consistent inclusion of the scale
parameter $\mu$ in the zeta function associated to the zero-point energy, as
done, usually, for the effective potential, leads, not to 
Myers' term, but to a null ZPA.

\section{Zero-Point Ambiguity}

In the functional approach to QFT, the vacuum energy density can be found by evaluating the minimum of the effective potential [1-7]. The energy density is a loop expansion (or equivalently in powers
 of $\hbar$). Roughly, it is the minimum of the classical potential plus quantum corrections, as is well known. 

For the sake of simplicity let $\phi(x)$ be a single real scalar field in an N-dimensional Minkowski
space-time, subject to the potential $V(\phi)$. The minimum of the effective potential, to first order in the loop expansion (or equivalently
in powers of $\hbar$), is given by[1,2]
\begin{eqnarray}
V_{ef}(\bar{\phi}) = V_{cl}(\bar{\phi}) + \frac{1}{2}
\frac{\hbar}{\Omega}\ln\det\biggl[
\frac{\delta^{2}S[\bar{\phi}]}{\delta\phi(x)\delta\phi(y)}\biggr] =
V_{cl}(\bar{\phi}) + V_{ef}^{(1)}(\bar{\phi}),
\end{eqnarray}
where $\bar{\phi}=<\phi>$ is the classical field, $S[\phi]$ is the
classical action, $\Omega=L^{m}T$ is the volume of the background
space-time manifold. Also $N= m + 1$, where $m$ is the number of spatial dimensions and in the classical potential $V_{cl}(\phi)$ mass and self-interactions terms 
are included.

 Making the usual analytic continuation to Euclidean space-time [2,4], the classical action can be written as
\begin{eqnarray}
S[\phi]= \int d^{N}x\biggl[\frac{1}{2}\partial_{\mu}\phi\partial_{\mu}\phi
+ V_{cl}(\phi)\biggr],
\end{eqnarray}
where an euclidean summation convention is understood for repeated
indexes. From (2) we get the matrix ${\it m}(x,y)$ of the quadratic variation of the action $S[\phi]$
\begin{eqnarray}
{\it m}(x,y) \equiv \frac{\delta^{2}S[\bar{\phi}]}{\delta\phi(x)
\delta\phi(y)} = \delta^{4}(x-y)[-\delta^{\mu\nu}\partial_{\mu}
\partial_{\nu} + V_{cl}^{\prime\prime}(\bar{\phi})].
\end{eqnarray}
Now, {\it m} is a real, elliptic and self-adjoint operator (because of the Euclidean analytic continuation) and for these kinds of operators we can 
define the so-called generalized zeta function. Let $\{\lambda_{i}\}$ be the eigenvalues of the operator $m(x,y)$. The generalized zeta function associated to ${\it M}(x,y)$ $(m \rightarrow {\it M} = \frac{m}{2\pi\mu^{2}})$ is defined by 
\begin{eqnarray}
\zeta_{\it M}(s) = \sum_{i}\biggl(\frac{\lambda_{i}}{2\pi\mu^{2}}\biggr)^{-s},
\end{eqnarray}
where we have introduced a unknown scale parameter $\mu$, with the dimensions of (length)$^{-1}$ or mass, in order to keep the zeta function dimensionless for all s.

Using the well-known relation [2,8]
\begin{eqnarray}
\ln\det{\it M} = - \frac{d\zeta_{\it M}(0)}{ds} \quad ,
\end{eqnarray}
the effective potential, to first order in the loop expansion, can be written
\begin{eqnarray}
V_{ef}^{(1)}(\bar{\phi}) = -\frac{1}{2}\frac{\hbar}{\Omega}
\frac{d\zeta_{\it M}(0)}{ds} \quad .
\end{eqnarray}

For most usual cases we can write the eigenvalues of the operator $m(x,y)$ as
\begin{eqnarray}
\lambda_{i,\omega} = \omega^{2} + h_{i}^{2} \quad ,
\end{eqnarray}
where $h_{i}$ are eigenvalues of the Hamiltonian operator $H$ and $\omega$ is a continuous parameter labeling the temporal part of the eigenvalues of the operator $m(x,y)$. This is a important hypothesis for our derivation as was
 also observed by Myers[10].

The generalized zeta function associated to the operator ${\it M}(x,y)$, defined by (4), can be written, using (7), as
\begin{eqnarray}
\zeta_{M}(s) = \int_{-\infty}^{\infty} \frac{d\omega}{2\pi}\sum_{i}\biggl[\frac{\omega^{2}}{2\pi\mu^{2}} + \frac{h_{i}^{2}}{2\pi\mu^{2}}\biggr]^{-s}T.
\end{eqnarray}
Using the relation[10]
\begin{eqnarray}
\int_{-\infty}^{\infty}\biggl(k^{2} + A^{2}\biggr)^{-s}d^{m}k =
\frac{\pi^{\frac{m}{2}}\Gamma(s-m/2)}{\Gamma(s)}\biggl(A^{2}
\biggr)^{\frac{m}{2}-s},
\end{eqnarray}
we can perform the above integral in $d\omega$ and get
\begin{eqnarray}
\zeta_{M}(s) = \frac{1}{2\sqrt{\pi}}\frac{\Gamma[s-1/2]}{\Gamma[s]}\zeta_{H}(s-1/2)T,
\end{eqnarray}
where $\zeta_{H}(s-1/2)$ is the generalized zeta function associated to the Hamiltonian operator $H$ and defined by
\begin{eqnarray}
\zeta_{H}(s-1/2) =  \sum_{i}\biggl(\frac{h_{i}^{2}}{2\pi\mu^{2}}\biggr)^{1/2-s}.
\end{eqnarray}
Another usual definition of vacuum energy (more intuitive than the minimum of the effective potential) is the sum over eigenvalues of the zero-point energy, namely
\begin{eqnarray}
\epsilon = \frac{\hbar}{2L^{m}}\sqrt{2\pi\mu^{2}}\sum_{i}\biggl( \frac{h^{2}_{i}}{2\pi\mu^{2}} \biggr)^{1/2}.
\end{eqnarray}

Using the generalized zeta function defined in (11) we can rewrite (12) as
\begin{eqnarray}
\epsilon = \frac{\hbar}{2L^{m}}\sqrt{2\pi\mu^{2}}\lim_{s\rightarrow 0}[\zeta_{H}(s-1/2)].
\end{eqnarray}
The inclusion of scale parameter $\mu$ in (12) and (13) is an important step since now 
from (6), (10) and (13) we find the relation between the effective potential and
 the zero-point energy
\begin{eqnarray}
V_{ef}(\bar{\phi}) = \epsilon + ZPA ,
\end{eqnarray}
where
\begin{eqnarray}
ZPA = \frac{\hbar}{2L^{m}}\sqrt{2\pi\mu^{2}}\lim_{s\rightarrow 0}\biggl[\psi(s-1/2)\frac{\zeta_{H}(s-1/2)}{\Gamma[s]} + \frac{\zeta_{H}^{\prime}(s-1/2)}{\Gamma[s]}\biggr] .
\end{eqnarray}

The above equation shows that the effective potential, in principle, can be different from the zero-point energy. This additional term we named {\it Zero-Point Ambiguity} (ZPA). It is, as written in (15), a non-renormalized
quantity.
In order to show that the renormalized ZPA vanishes, we can write  $\zeta_{H}(s-1/2)$ from (10) as
\begin{eqnarray}
\zeta_{H}(s-1/2) = g(s)\Gamma[s] ,
\end{eqnarray}
where g(s) is analytic at $s=0$, since $\zeta_{M}(s)$ is analytic at $s=0$.

Substituting the (16) in (15) we obtain for the ZPA
\begin{eqnarray}
ZPA = \frac{\hbar}{2L^{m}}\sqrt{2\pi\mu^{2}}\lim_{s\rightarrow 0}\biggl[\psi(s-1/2)g(s) + g^{\prime}(s) +  \frac{\psi(s)}{\Gamma[s]}g(s)\Gamma[s]\biggr] .
\end{eqnarray}

A quick glance of the above equation shows that it has a simple pole at $s=0$ in the third term, since $\Gamma(s)$ has the asymptotic behavior
\begin{eqnarray}
\frac{2\sqrt{\pi}\Gamma[s]}{\Gamma[s-1/2]} = -\frac{1}{s} .
\end{eqnarray}

In order to extract the finite part from equation (17) we use the 
Salam-Strahdee prescription, that is, we multiply the third term by $s$ and compute the derivative with respect to $s$ at $s=0$ [9]. We get
\begin{eqnarray}
\lim_{s\rightarrow 0}\biggl[\frac{\psi(s)}{\Gamma[s]}g(s)\Gamma[s]\biggr] =  -\psi(-1/2)g(0) - g^{\prime}(0) .
\end{eqnarray}
Now, from (17) and (19), we easily get the renormalized ZPA, which turns out to be
\begin{eqnarray}
ZPA^{R} = 0 .
\end{eqnarray}
From (14) and (20) we obtain our final result 
\begin{eqnarray}
V^{R}_{ef} = \epsilon^{R}.
\end{eqnarray}
Observe that the Salam-Strathdee prescription has provided the necessary analytic continuation for the ZPA
such that the renormalized ZPA vanishes.

\section{Myers' Extra Term and ZPA}

In his paper Myers [10] stressed that the vacuum energy density calculated via an effective potential could be interpreted as the sum over zero-point energies except for an extra term. When this extra term is zero, both definitions of vacuum energy coincide. Now a question  arises: Is the ZPA the Myers' extra term? 
The answer is no. Myers' extra term arises because the scale parameter $\mu$ has not been included in the definition of the generalized zeta function associated to the Hamiltonian operator for the sum over eingenvalues of zero-point energy.
As stressed above, this was already noted by Treml[11]. As we will soon see, even after  renormalization, Myers' extra term  does not vanish. On the other hand, the ZPA arises when we compare two terms,namely  $V^{(1)}_{ef}$ given by (6) and  $\epsilon$ given by (13). The evaluation of $V^{(1)}_{ef}$, using the zeta function method implicitly
removes the divergences while for $\epsilon$ we must run over an explicit renormalization procedure in order to eliminate the divergences. The 
Salam-Strathdee prescription provides such a  
renormalization scheme and we get a vanishing renormalized ZPA as shown in the section 2. The details are as follows below.

As in section 2, we define the difference between the effective potential 
and zero-point energy as
\begin{eqnarray}
\delta\epsilon = V^{(1)}_{ef} - \bar{\epsilon},
\end{eqnarray}
where $\bar{\epsilon}$ is defined as
\begin{eqnarray}
\bar{\epsilon} = \frac{\hbar}{2L^{m}}\lim_{s\rightarrow 0}[\zeta_{h}(s-1/2)],
\end{eqnarray}
with
\begin{eqnarray}
\zeta_{h}(s-1/2) =  \sum_{i}\biggl(h_{i}^{2}\biggr)^{1/2-s}.
\end{eqnarray}
Now, note that, in the above definition, we did not include any scale
 parameter $\mu$ for the zero-point energy, whereas, for the effective potential
 defined by (4), we have performed a scaling transformation on the operator $m$ in order to keep the zeta function dimensionless for all $s$. Now, it is well-known that the effective potential in (6) can be written as [8]
\begin{eqnarray}
V^{(1)}_{ef}(\bar{\phi}) = - \frac{\hbar}{2\Omega} \zeta^{\prime}_{\it M}(0) = - \frac{\hbar}{2\Omega} \zeta^{\prime}_{m}(0) - \frac{\hbar}{2\Omega}\zeta_{m}(0)\ln(2\pi\mu^{2}),
\end{eqnarray}
with $\zeta_{m}(s)$ defined as
\begin{eqnarray}
\zeta_{m}(s) = \sum_{i}\lambda_{i}^{-s}.
\end{eqnarray}
According to (10) we can write
\begin{eqnarray}
\zeta_{m}(s) = \frac{1}{2\sqrt{\pi}}\frac{\Gamma[s-1/2]}{\Gamma[s]}\zeta_{h}(s-1/2)T \quad .
\end{eqnarray}
Using relation (18), the above equation can be rewritten as
\begin{eqnarray}
\zeta_{h}(s-1/2) =  - \frac{1}{s}\zeta_{m}(s)\frac{1}{T}.
\end{eqnarray}
Using  (23) and (25) in (22) we find
\begin{eqnarray}
\delta\epsilon = - \frac{\hbar}{2\Omega}\zeta^{\prime}_{m}(0) - \frac{\hbar}{2\Omega}\zeta_{m}(0)\ln(2\pi\mu^{2}) - \frac{\hbar}{2L^{m}}\lim_{s\rightarrow 0}[\zeta_{h}(s-1/2)].
\end{eqnarray}
Note that the last term is divergent since it has a simple pole at $s=0$, as 
is easily seen from (28). Then, the Myers' non-renormalized extra term reads
\begin{eqnarray}
\delta\epsilon = - \frac{\hbar}{2\Omega}\zeta^{\prime}_{m}(0) - \frac{\hbar}{2\Omega}\zeta_{m}(0)\ln(2\pi\mu^{2}) + \frac{\hbar}{2L^{m}}\lim_{s\rightarrow 0}[\frac{1}{s}\zeta_{m}(s)].
\end{eqnarray}
In order to renormalize it  we use, again, the Salam-Strathdee prescription. We obtain, easily,
\begin{eqnarray}
\delta\epsilon = - \frac{\hbar}{2\Omega}\zeta_{m}(0)\ln(2\pi\mu^{2}),
\end{eqnarray}
which is Myers' extra term[10]. So, after renormalization, Myers'
extra term does not vanish and, as Myers observe, only when $\zeta_{m}(0)=0$
the effective potential coincides with the zero-point energy. But this is not 
correct since we must include the scale parameter also in the zeta function
associated to the zero-point energy as we have done in the effective potential in order to keep it dimensionless. Including the scale parameter we obtain our
null ZPA. 

To show, explicitly, the relation between the ZPA, (15) and the Myers extra term, (31), we note that from (11) and (24) we have the relation
\begin{eqnarray}
\zeta_{H}(s-1/2) = (2\pi\mu^{2})^{s-1/2}\zeta_{h}(s-1/2).
\end{eqnarray}
Substituting (32) in (15) we get (before taking the limit $s \rightarrow 0$)
\begin{eqnarray}
ZPA(s) = \frac{\hbar}{2\Omega} \frac{2\sqrt{\pi}}{\Gamma(s-1/2)}(2\pi\mu^{2})^{s} \biggl[ \zeta_{m}(s)\ln(2\pi\mu^{2}) + \zeta^{\prime}_{m}(s) \biggr] + \nonumber \\
+ \frac{\hbar}{2\Omega}(2\pi\mu^{2})^{s} \frac{\psi(s)}{\Gamma(s)} \biggl( \frac{-1}{s} \biggr) \zeta_{m}(s).
\end{eqnarray}
Note that the first term on the right-hand side of above equation is the
 Myers' term (when $s=0$) and comes from the derivative of $\zeta_{H}(s-1/2)$. Note also that this term is cancelled, after renormalization of the last term.  In other words, ZPA is a well defined concept, while Myers' term
is not and, in fact, it does not exist if a scale parameter $\mu$ is consistently
introduced.

If we fixed the scale parameter, ($2 \pi \mu = 1 $), for example, we still can 
define a ZPA by using
\begin{eqnarray}
V_{ef} = - \frac{\hbar}{2\Omega}\zeta^{\prime}_{m}(0),
\end{eqnarray}
and
\begin{eqnarray}
\epsilon = \frac{\hbar}{2\Omega}\lim_{s\rightarrow 0}[\zeta_{h}(s-1/2)] .
\end{eqnarray}

The ZPA, in this case, is given by 
\begin{eqnarray}
ZPA = \frac{\hbar}{2L^{m}} \lim_{s\rightarrow 0}\biggl[\psi(s-1/2)\frac{\zeta_{h}(s-1/2)}{\Gamma[s]} + \frac{\zeta_{h}^{\prime}(s-1/2)}{\Gamma[s]}\biggr] .
\end{eqnarray}

\section{Conclusion}

We have shown that, in Minkowski space-time, for a large class of geometries 
and for a general potential, the effective potential is equivalent to the sum 
over zero-point energy. It is important to note that, for flat space-times,
the renormalization prescriptions above are well-defined, whereas for curved
space-times new ambiguities may appear[15]. Also it would be interesting to know whether
the analysis above can be worked out, using other methods of regularization, such as dimensional regularization, for example.

In considering Myers' extra term, we recall that the correct procedure is to introduce a scale parameter $\mu$ into any zeta function in order to keep it dimensioless for all $s$. Therefore, we should not  consider Myers' extra term as a real difference between the effective potential and zero-point energy.

We point out that the result of the equivalence between effective potential and zero-point energy in this work is more general than our earlier result [12] because it is valid for a large class of geometries (submanifolds) in
Minkowski space-time as well for a general potential, so that no explicit
model must be considered. Also it is easy to see that our methods are valid
for an arbitrary dimension of the space-time. Finally, it should be noted that our result is valid for one loop (see eq(s) (6) and (12)). The result for higher order terms
in the loop expansion deserves a further study.

\section{Acknowledgments}
A. Maia thanks Robert Brandenberger, Jose A. S. de Lima and Joao Nunes for valuable suggestions and the Physics Department at Brown
University for the hospitality and also FAPESP (Sao Paulo Research Foundation-
Brazil) for a Postdoctoral Fellowship. This work was partially supported by
FAPESP and by U.S. Department of Energy under Grant No DE-FG02-91ER-40688, Task 
A.

\end{document}